# IM NORMAE: THE DEATH SPIRAL OF A CATACLYSMIC VARIABLE?

Joseph Patterson,[1] Jonathan Kemp,[2] Berto Monard,[3] Gordon Myers,[4] Enrique de Miguel,[5,6]

Franz-Josef Hambsch,[7] Paul Warhurst,[8] Robert Rea,[9] Shawn Dvorak,[10] Kenneth Menzies,[11]

Tonny Vanmunster,[12,13] George Roberts,[14] Tut Campbell,[15] Donn Starkey,[16] Joseph Ulowetz,[17]

John Rock,[18] Jim Seargeant,[19] James Boardman,[20] Damien Lemay,[21] David Cejudo,[22]

& Christian Knigge[23]

[1] Department of Astronomy, Columbia University, 550 West 120th Street, New York, NY 10027; jop@astro.columbia.edu; 0000-0003-2762-5572
[2] Mittelman Observatory, Middlebury College, Middlebury, VT 05753; jkemp@middlebury.edu; 0000-0002-8675-8079
[3] CBA–Kleinkaroo, Klein Karoo Observatory, PO Box 281, Calitzdorp 6660, South Africa; astroberto13m@gmail.com
[4] CBA–San Mateo, 5 Inverness Way, Hillsborough, CA 94010; gordonmyers@hotmail.com; 0000-0002-9810-0506
[5] Departamento de Física Aplicada, Facultad de Ciencias Experimentales, Universidad de Huelva; 0000-0002-1381-8843
[6] CBA–Huelva, Observatorio del CIECEM, Parque Dunar, Matalascañas, 21760 Almonte, Huelva, Spain; edmiguel63@gmail.com
[7] CBA–Mol, ROAD Observatory, Oude Bleken 12, B-2400 Mol, Belgium; hambsch@telenet.be
[8] Department of Physics, University of Auckland, Private Bag 92019, Auckland, New Zealand; p.warhurst@auckland.ac.nz
[9] CBA–Nelson, Regent Lane Observatory, 8 Regent Lane, Richmond, Nelson 7020, New Zealand; reamarsh@slingshot.co.nz
[10] CBA–Orlando, Rolling Hills Observatory, 1643 Nightfall Drive, Clermont, FL; sdvorak@rollinghillsobs.org
[11] CBA–Framingham, 318A Potter Road, Framingham, MA 01701; kenmenstar@gmail.com
[12] CBA–Belgium, Walhostraat 1A, B-3401 Landen, Belgium; tonny.vanmunster@gmail.com
[13] CBA–Extremadura, e-EyE Astronomical Complex, ES-06340 Fregenal de la Sierra, Spain
[14] CBA–Tennessee, 2007 Cedarmont Drive, Franklin, TN 37067; georgeroberts0804@att.net
[15] CBA–Arkansas, 7021 Whispering Pine, Harrison, AR 72601; jmontecamp@yahoo.com
[16] CBA–Indiana, DeKalb Observatory (H63), 2507 County Road 60, Auburn, IN 46706; donn@starkey.ws
[17] CBA–Illinois, Northbrook Meadow Observatory, 855 Fair Lane, Northbrook, IL 60062; joe700a@gmail.com
[18] CBA-Wilts, 2 Spa Close, Highworth, Swindon, Wilts SN6 7PJ, United Kingdom; john@highworthobs.fsnet.co.uk
[19] CBA–Edgewood, 11 Hilltop Road, Edgewood, NM 87015; jimsarge@gmail.com
[20] Krazy Kritters Observatory, 65027 Howath Rd, De Soto, WI, 54624, USA; j_boardmanjr@yahoo.com
[21] CBA-Quebec, 195 Rang 4 Ouest, St-Anaclet, Quebec G0K 1H0, Canada; damien.lemay@globetrotter.net
[22] CBA–Madrid, Observatorio El Gallinero, El Berrueco, 28192 Madrid, Spain; davcejudo@gmail.com
[23] School of Physics and Astronomy, University of Southampton, Highfield, Southampton, SO17 1BJ, United Kingdom; c.knigge@soton.ac.uk; 0000-0002-1116-2553

**ABSTRACT**

We present a study of the orbital light curves of the recurrent nova IM Normae since its 2002 outburst. The broad "eclipses" recur with a 2.46 hour period, which increases on a timescale of $1.28(16)\times10^6$ years. Under the assumption of conservative mass-transfer, this suggests a rate near $10^{-7}$ $M_\odot$/year, and this agrees with the estimated *accretion* rate of the postnova, based on our estimate of luminosity. IM Nor appears to be a close match to the famous recurrent nova T Pyxidis. Both stars appear to have very high accretion rates, sufficient to drive the recurrent-nova events. Both have quiescent light curves which suggest strong heating of the low-mass secondary, and very wide orbital minima which suggest obscuration of a large "corona" around the primary. And both have very rapid orbital period increases, as expected from a short-period binary with high mass transfer from the low-mass component. These two stars may represent a final stage of nova — and cataclysmic-variable — evolution, in which irradiation-driven winds drive a high rate of mass transfer, thereby evaporating the donor star in a paroxysm of nova outbursts.

**Concepts**

Cataclysmic variable stars (203), Classical novae (251), Close binary stars (254), Interacting binary stars (801), Novae (1127), Recurrent novae (1366), Stellar accretion (1578), Stellar accretion disks (1579)

**Objects**

V* AA Dor, V* IM Nor, V* T Pyx, V* QR And, V* V Sge, V* V617 Sgr

# 1 INTRODUCTION

## 1.1 The T Pyx Mystery

The recurrent nova T Pyxidis erupted for the sixth time in 2011, bringing that famous star yet again to the world's attention. Many observing campaigns were carried out, revealing some fascinating results: a "supersoft" X-ray source (Chomiuk et al. 2014), an expanding radio nebula (Nelson et al. 2014), improved measures of the distance (Sokoloski et al. 2013, Schaefer 2018), and a discrete orbital-period change across the outburst (Patterson et al. 2017, hereafter P17). The new distance, dominated by the *Gaia* parallax, yields an improved estimate of the star's luminosity ($M_V$ = +2 at quiescence, and −8 at the peak of eruption). Maximum light is no surprise, since all novae climb to the Eddington limit at maximum. But the quiescent Mv is a shocker. Classical novae at quiescence are typically near $M_V$ = 4.5, so the quiescent T Pyx is ~2.5 magnitudes "too bright".

Even more puzzling, T Pyx's 1.8 hour orbital period is in the domain where nearly all cataclysmic variables (CVs) have $<M_V>$ in the range +8 to +11 (Figures 5 and 7 of Patterson 2011, hereafter P11). So the "quiescent" T Pyx is ~8 magnitudes too bright for its natural community of stars. This huge discrepancy is particularly flagrant, because for CVs of similar orbital period, present-day theory anoints gravitational radiation — which operates at a known rate and produces roughly $M_V$ ~ +10 — as the sole driver of mass transfer and evolution.

Assuming a CV space density of ~$10^{-5}$/pc$^3$ (P11; Patterson 1984, hereafter P84; Pretorius & Knigge 2012; Pala et al. 2020), there should be about a few hundred thousand CVs out to the distance of T Pyx. Most are probably of short $P_{orb}$, like ~80% of CVs in the solar neighborhood. A slow nova like T Pyx, which erupts to 7th magnitude every 25 years, should be an easy target for nova hunters and all-sky patrols. Nearby members of the class would even be visible to the naked eye. And yet, 130 years and six eruptions later, there is still only one T Pyx.

That's the mystery. It's "one in a million".

## 1.2 IM Normae, a Possible Comrade

IM Normae was discovered by I. Woods in 1920 as a 9th magnitude star on Harvard photographic plates, but the quiescent counterpart eluded detection (Bailey 1920, Elliot and Liller 1972, Wyckoff & Wehinger 1979). It remained simply a good nova candidate, about which practically nothing was known, until a second outburst was detected (Liller 2002). The visual record then established that it was a slow nova, resembling T Pyx, with *V* = 7.8 at maximum and an 18th magnitude prenova counterpart (Retter, O'Toole, & Starrfield 2002; Kato, Yamaoka, Liller, & Monard 2002). Figures 5 and 6 of Strope et al (hereafter SSH) show the eruption light curves of the two stars, warranting the description "resembling T Pyx".

One year after eruption, Woudt & Warner (2003) obtained time-series photometry of the postnova, and found a stable and smooth variation with a period of 2.46 hours. This was reminiscent of T Pyx, the only previously known recurrent nova of short orbital period. The orbital light curves of the two stars, discussed below, are very similar. This particular shape is quite unusual for a CV — giving extra credence to the hypothesis that the two stars may share a common underlying physics.

Two stars do not a class make, and do not necessarily explain, clarify, or dismiss the mystery of T Pyx. But you have to start somewhere. So we undertook a long-term campaign of time-series photometry, which is reported here.

## 2 THE OBSERVATIONS

The new IM Nor observations consist of time-series photometry during 2003−2020, with several telescopes in the Center for Backyard Astrophysics network (CBA, Patterson et al. 2013). We accumulated ~1000 hours of observation, spread over 210 nights. Much of the early observing was done at Monard's two South Africa stations, CBA-Pretoria (2002−2009) and CBA-Kleinkaroo (2010−2020). Each used a Meade 30 cm Schmidt-Cassegrain telescope equipped with SBIG ST7 and ST8-XME CCD cameras. Most of the later photometry used Myers' 42 cm and 50 cm ODK telescopes at Siding Spring Observatory. To maximize signal-to-noise on this faint star, we always observed with a clear or clear-with-UV-cutoff filter. For blue stars like cataclysmic variables, this white-light photometry produces an effective wavelength near 6000 Å ("pink"). A few simultaneous measures of $V$ and "clear" magnitudes established a rough calibration.[24] During the 2003−2020 span of our observations, the star's mean brightness declined smoothly from $V$ = 17.4 to 18.6.

Midway through this campaign, as we became increasingly aware of IM Nor's high "quiescent" luminosity and other resemblances to T Pyx, we began observing two other short-period, very luminous binaries: the supersoft X-ray source QR Andromedae, and V617 Sagittarii, presently classified as a :V Sge star (there are grounds for considering these to belong also in the supersoft class). These observations will be described in § 7.

## 3 THE PERIODIC SIGNAL, ITS WAVEFORM, AND THE RATE OF PERIOD CHANGE

All the IM Nor light curves resemble that shown by Woudt & Warner (2003), with a broad primary eclipse, and a small secondary dip. We measured the light curves for times of primary eclipse, where each timing represents 1−6 orbits, depending on data quality. These 53 times of minimum light are listed in Table 1.

The top frames of Figure 1 show the mean light curve during year 1 and year 15 after the eruption. The waveform changes smoothly from one to the other during the intervening years (*viz.*, both eclipse depths increasing with time). For comparison, the lower frames of Figure 1 show the orbital light curves of T Pyx in year 1 and year 6 after the 2011 eruption. Figure 4 of P17 also illustrates for T Pyx the pattern of orbital-waveform change: both eclipse depths increase as the nova declines (although in the customary language, all these light curves are considered "at quiescence").

---

[24] Because of their intrinsically blue color, most CVs have "clear" magnitudes ~0.10−0.15 mag too faint, relative to a true V magnitude. However, this correction was found to be very small for IM Nor, a highly reddened star.

Why do we call the primary minimum an *eclipse*? At first sight, it seems far too broad to be an actual geometrical eclipse. In the well-studied CVs, the broadest obvious eclipses range from orbital phase 0.89 to 1.11; and those deep-eclipsing stars (DQ Her and BT Mon) are thought to be very close to edge-on ($i > 85°$). But the primary minima in IM Nor and T Pyx are ~0.4 cycles wide, and yet the shallow depths of the minima suggest modest inclinations. Nevertheless, in T Pyx the radial-velocity and photometric periods are identical, and minimum light is known to occur at superior conjunction of the emission-line source, presumed to be the disk (Uthas, Knigge & Steeghs 2010; hereafter UKS). This is also true for the only other recurrent nova sometimes described as "short-period" (CI Aqulae, $P_{orb}$ = 15 hr; Wilson & Honeycutt 2014, Schaefer 2011). Thus, radial-velocity phasing suggests that the primary minimum really is an eclipse, although of a sort not commonly found in the ranks of CVs. It is less clear that the secondary dip is a true eclipse. We discuss these matters further in Sec. 7.

From the IM Nor eclipse times we derive an approximate period of ~0.10263317 d, and Figure 2 shows an O−C diagram with respect to that period. The upward curvature signifies an ever-increasing period, and the points are well fit by the ephemeris[25]

$$\text{Min. light} = \text{HJD } 2{,}452{,}696.5265(2) + 0.1026327(1)\ E + 1.12(7) \times 10^{-12}\ E^2. \quad (1)$$

This ephemeris corresponds to the parabolic fit in Figure 2, and indicates period increase at a rate

$$P/\dot{P} = 1.28(16) \times 10^6 \text{ years}. \quad (2)$$

Except for T Pyx and the supersoft binaries, this is much (>10×) faster than any other CV previously studied.

Small "orbital period changes" are fairly common in CVs, but are usually not susceptible to direct interpretation as an effect of evolution or mass transfer, because the apparent period changes are commonly of both signs (+ and −). Also, they're very small: the amplitude of wiggles in the O−C usually doesn't exceed 0.005 cycles — ten times smaller than the effect seen here, and 100x smaller than the effect seen in T Pyx. The subtleties of these effects, and difficulties in interpreting them, have been extensively discussed (Pringle 1975, Warner 1988, Applegate 1989, Marsh & Pringle 1990, Han et al. 2016), but remain mostly unexplained. T Pyx is different. The size of the wiggles and speed of the long-term $\dot{P}$ are ~100× greater, the direction of period change is monotonic (P increasing), and the radial velocities precisely track the variable-period photometric ephemeris (UKS). These strongly warrant an orbital-period interpretation. IM Nor, definitely a comrade in eruption type, has a similar light curve and a similar $\dot{P}$. The light curve also resembles the known-to-be-orbital light curves of some supersoft binaries (*e.g.*, CAL 87, see Figure 4 of Schandl et al. 1997; some of these light curves and period changes are discussed below).

---

[25] Nearly all data collected and discussed in this paper were obtained with small telescopes operated by "amateur" astronomers, and consistently reported with JD or HJD dates. The HJD correction is large, so we have adopted that as our "sufficiently uniform" time base. The barycentric correction is completely negligible, so we ignore it. For long baselines, the correction to ephemeris time (essentially, leap seconds) is not negligible, but is quite small, since leap seconds have been rare in recent years. With such a wide range of observers and at-the-telescope software, we choose to minimize ambiguity and *ignore the ET correction*. But as the time baseline grows longer, users of this data should probably apply it.

Thus all the evidence points to an orbital-period interpretation for IM Nor. We shall also assume that the orbital period is actually increasing at this rate (and not plagued by a phase shift which mysteriously and sadistically wiggles back and forth with very large amplitude on a timescale much longer than our 17-year baseline).

## 4 THE NOVA RECURRENCE PERIOD

The recurrence period for nova eruptions is nominally 82 years, but this should be considered an upper limit. The identification and precise position of the star was not even known until the 2002 eruption; targeted searches before then must have been rare (and ambiguous). Serendipitous discovery of an eruption is possible, but the far-southern location and crowded field take the star away from the usual hunting grounds of visual nova observers. It's entirely possible that some outbursts have been missed[26] since 1920, and hence we consider the outburst period to be 82/*N* years, where *N* could be 1, 2, 3, or maybe even 4.

## 5 ACCRETION RATE, DISTANCE, LUMINOSITY

In a binary, a natural mechanism for period increase is mass transfer from the low-mass star to the high-mass star. And this is what CVs do for a living, since they are normally powered by accretion. Assuming[27] conservative mass transfer (total mass and angular momentum conserved), the mass transfer rate is given by

$$\dot{M} = M_1 M_2 / [(3P/\dot{P})(M_1 - M_2)]. \quad (3)$$

At $P_{orb}$ = 2.46 hr, IM Nor is quite near the middle of the famous 2−3 hour "period gap". According to theory (and to a lesser extent, observation), the secondaries of such stars should be near 0.2 $M_\odot$ [for a thorough treatment of this, see Table 3 of Knigge (2006)]. Since IM Nor is also a recurrent nova, its white dwarf is probably fairly massive. Assuming $M_1$ = 1.0 $M_\odot$, Eqs. (2) and (3) then imply $\dot{M}$ = 6.6×10$^{-8}$ $M_\odot$/yr. This is a very high accretion rate for a CV, although 3× less than the corresponding value for T Pyx.

---

[26] As also argued by S10. Even the original 1920 outburst was, to some extent, missed. It was discovered 50 years later from archival plates (Eliot & Liller 1972). And the 2002 outburst was only noticed after Wyckoff & Wehinger (1979) and Duerbeck (1987) specifically called attention to the still-unsuccessful identification of the star which erupted in 1920. Kato et al (2002) describes the history.

[27] A merely default assumption. Nova events imply mass and angular-momentum loss, and recurrent novae spend a lot of time in eruption. Even after eruption, some mass and angular-momentum loss may continue (thus increasing and decreasing *P*, respectively). These issues will be revisited in § 7.

A first test for reasonableness is: does this accretion rate produce recurrent-nova events every 80 years? The answer appears to be **yes**. In table 3 of Yaron et al. (2005), an accretion rate of $10^{-7}$ $M_\odot$/yr produced nova outbursts in less than 100 yr for all white dwarfs of 1 $M_\odot$ or greater.

A second test is: can the quiescent luminosity be produced by accretion? This one is tougher, because we really don't know the distance to IM Nor. The strong interstellar absorption lines favor $d$ >2.5 kpc (Duerbeck et al. 2003), and the limit on the Gaia EDR3 parallax does not improve on this.

Since the nova light curve resembled that of T Pyx, we could use that as a model. The eruption light curves of both stars are presented in Figures 2−7 of S10. IM Nor is fainter by ~2.3 mag in eruption and ~2.8 mag in quiescence, suggesting that it might be ~3.2× more distant. On the other hand, IM Nor is much redder — by Δ(B−V) = 0.9 according to Table 30 of S10. IM Nor's location just 2.5° from the Galactic plane (versus 10° for T Pyx) also suggests a large reddening. If we adopt the Galaxy-wide average of $A_V$ = 3.1 E(B−V), then that extra reddening corresponds to an extra $A_V$ of 2.8 — suggesting that T Pyx and IM Nor are actually at a similar distance. A similar estimate, based on different arguments, was made by S10. For T Pyx, the *Gaia EDR3* parallax gives the best distance estimate (2900±300 pc). For simplicity we will adopt $d$ = 3000 pc for both stars. The above "intrinsically like T Pyx" argument would suggest a total $A_V$ = 2.8 + 1.1 (the P17 estimate for T Pyx) = 3.9. But Ozdonmez et al. (2018) estimates E(B−V) = 0.8±0.2, and our unpublished fit to the UV spectrum suggests E(B−V) = 1.0±0.3. So these various rough estimates suggest $A_V$ ~3.0. With our assumed 3 Kpc distance and an observed mean brightness of V = 18.5, we then obtain a dereddened $M_V$ = 3.1. The eclipses in IM Nor are fairly deep, and we adopt an additional correction of 0.3 mag to convert it to an average inclination (57°). So we credit IM Nor with $M_V$ = 2.8.

In so compact a binary, this requires a very hot object, and we estimate a bolometric correction near −3.0. Adopting $M_{bol}$ = −0.2 and assuming the light comes from accretion through a disk, we then use

$$L = GM\dot{M}/2R \quad (4)$$

to obtain $\dot{M}$ = 5x$10^{-8}$ $M_\odot$/yr. Within the uncertainties in this estimate, we conclude that the answer to the second test is probably **yes**: an accretion rate near $10^{-7}$ $M_\odot$/year is consistent with our estimate of quiescent luminosity, as well as the orbital-period change, plus the theoretical interpretation of the recurrence time (requiring a fairly massive WD accreting at a fairly high rate).

## 6 IM NOR AMONG THE CVS

The "comrade to T Pyx" hypothesis has survived these tests. The stars show similar eruption light curves, orbital light curves in quiescence, changes in orbital light curve during decline, and orbital periods.  And both seem to be transferring matter at a rate[28] near $10^{-7}$ $M_\odot$/yr — sufficient to power a quiescent accretion luminosity of ~$10^{36}$ erg/s and drive a very rapid period change (evaporating the secondary in ~$10^6$ years, assuming the last century to be representative).

This luminosity, and corresponding accretion rate, is remarkable for stars of such short $P_{orb}$.  For disk-accreting white dwarfs, time-averaged $M_V$ is a pretty good proxy for $\dot{M}$, and is available for all CVs with long-term light curves (such as those archived by the AAVSO).  Figure 3 illustrates the place of IM Nor and T Pyx in the family of accretion-powered nonmagnetic CVs.  This is a greatly expanded version of Figure 7 in P84 (based on that paper's Table 1), and a slightly expanded version of Figure 5 in P11 (based on the electronic version of that paper's Table 2), and Figure 9 in P17.  Those papers describe in detail the methods for estimating $<m_V>$, distance[29], and binary inclination.  For the classical novae, the data are primarily drawn from the tabulations of Schaefer (2018), Warner (1987), and Duerbeck (1992).

We correct the various raw estimates of $<m_V>$ to a standard value of $i = 57°$, which is a theoretical average over 0−90°.  Since inclination is usually poorly known, we apply this correction [Eq. (2) of P11] only to obviously edge-on and face-on binaries.  We also subtract the WD's measured or estimated flux, which is usually negligible but matters slightly for the intrinsically faint stars.

Crosses are dwarf novae and novalike variables, and dots are old novae (20−150 years old, and therefore excluding the several very faint *pre-nova* measures by Schaefer & Collazzi 2010)  The general patterns reproduce those discussed in P84 and P11, but with IM Nor and T Pyx identified by name.  These two stars appear to be different.

Figure 3 contains all stars for which we judged the data to be adequate — usually requiring a baseline of at least 10−20 years of frequent observation (the same standard applied in the P84 and P11 studies).  But their populations in the figure certainly do not reflect their relative populations in the sky.  Most of the crosses are dwarf novae, with distances usually in the range 100−400 pc.  The nova remnants are more distant, typically ~1 kpc.  And T Pyx and IM Nor, the only known members of their class, are at ~3 kpc.  This underlines how rare these two stars are.  After counting all the known CVs (not just the ones with extensive long-term records, appearing in Figure 3) and accounting for their likely distances, we estimate that "T Pyx stars" are ~0.001% of all CVs.

---

[28]  For T Pyx, this matter is discussed thoroughly by Godon et al. (2018), mainly through analysis of the continuum fluxes.  Normally, period change would be a superior way to estimate d$M$/dt.  But the latter is clouded by uncertainties in mass loss and angular momentum loss.  Uncertainties in distance and reddening also cloud the estimates.  But these two stars are so unusual — so distinctive from other novae — that we lump them together and charge ahead.

[29]  Distances are based on a weighted average of trig parallax, photometric parallax of the secondary, photometric parallax of the WD, proper motion, strength of emission lines, interstellar reddening — in usually descending order of weight.  The tables in the original papers identify which clues are significant for which stars, but some of these weights (especially trig parallax) have changed over the years.  We have incorporated the new parallaxes from *Gaia*;  but surprisingly, they do not have much impact on this figure.  In part, this is because they are broadly consistent with previous distance estimates (including previous parallaxes).  More importantly, it is because the time-averaging of $<m_V>$ is the dominant uncertainty — since most of these stars are dwarf novae, which radiate most of their luminosity in outburst.  Long-term AAVSO records are especially critical, because they track recurrence periods and shapes of the decline. Trolling through these records is a labor well-suited for a pandemic.

According to the theory of CV evolution, $<\dot{M}>$ and therefore $<M_V>$ should be basically set by $P_{orb}$. [The WD mass can also be significant, but is usually of minor importance since WD masses are usually close to 0.8−0.9 $M_\odot$; see Figure 2 of Knigge (2006) and Figure 1 of Zorotovic et al. (2011)]. In the customary account of nova evolution, that well-populated lower curve is the main story of CV evolution below the period gap, but each star can experience a classical-nova eruption, thereby vaulting the star into the upper regions, where it stays for many years. *Something* keeps the accretion rate very high for at least a few hundred years. Hundreds or thousands[30] of years later, it settles back to near-quiescence, and the cycle repeats.

But some of these stars never get the opportunity to rest after their nova ordeals, because new classical-nova eruptions occur after a mere ~50 years. It's not clear if these stars — T Pyx and IM Nor — will *ever* get a chance to join their quiescent brethren in the general population at the lower regions of Figure 3.

All seven of the bright short-period stars are recent classical novae (T Pyx, IM Nor, GQ Mus, CP Pup, RW UMi, V1974 Cyg, and — arguably — BK Lyn). Except for T Pyx and IM Nor, these novae seem roughly consistent with their long-$P_{orb}$ cousins: eruption light curves not systematically different, not known to be recurrent, and clustering near $M_V$~+4 in quiescence. What makes T Pyx and IM Nor different from the other 5?

The recurrence timescale is sensitive to WD mass and accretion rate. In the vicinity of $M_1$ = 1.0 $M_\odot$, it scales roughly as $M_1^{-7}$ and $\dot{M}^{-1.3}$ in the models of Yaron et al. (2005). So a major factor could be simply the idiosyncrasy of WD mass.

But the deeper reason may lie in the donor star, since it powers the very high accretion rate. Actually, *all* the bright short-$P_{orb}$ stars in Figure 3 need explanation: their donors are absolutely not normal, because they are transferring mass much too fast.

This is an important point. These recent novae may well (and probably do) have hot WDs which heat the disk and thereby enhance $M_V$ over what it would be from the accretion rate alone. But it would be remarkable if such heating entirely mimicked all the effects of a high accretion rate: rapid flickering, quasi-periodic oscillations, weak accretion-disk lines, positive and negative superhumps, etc. (No single old nova has all these properties; but collectively, they are all present — and are the standard signatures of disk accretion in CVs.) We conclude that these puny secondaries (all <0.25 $M_\odot$ since they're all below the period gap) really are transferring matter at unnaturally high rates — unlike the situation at longer $P_{orb}$, where post-novae are roughly as bright as other CVs which have not had recorded nova eruptions. It's worth repeating: ***this needs explanation***.

---

[30] At least 3,000 years, according to the P13 scenario. This estimate is ~100× longer than a commonly quoted figure for classical novae generally. The reason is that most classical novae come from binaries of long $P_{orb}$, and such stars have powerful machines (called "magnetic braking", although their nature is not securely known) for stimulating mass transfer and thus generating luminosity unrelated to the nova event. Novae of **short** $P_{orb}$ are better tests for estimating the true timescale of decline — which must be at least many hundreds of years, since all such stars are still at least 3 magnitudes brighter than the great family of short-$P_{orb}$ stars in Figure 3 (the "CV main sequence").

# 7 HEATING EFFECTS

## 7.1 Heating and the Orbital Light Curve

The very high accretion rate in IM Nor and T Pyx is most naturally explained if the donor is strongly heated, possibly by radiation from the recently-erupted WD. The orbital light curve can then be a good diagnostic of this irradiation.

The natural result of strong irradiation is a quasi-sinusoidal "reflection effect", with maximum light at superior conjunction of the heated star. The textbook example of a reflection-effect light curve is AA Dor, a compact ($P_{orb}$ = 5.9 hour) and detached eclipsing binary, which contains a 42,000 K sdO star and a low-mass (0.08 $M_\odot$) secondary (Hilditch et al. 2003). Figure 4 shows the CBA orbital light curve of this star. Because the orbital inclination is very high, there are eclipses of both stars; and because there are no complications from mass-transfer, AA Dor offers a precise laboratory for measuring stellar parameters and understanding the atmospheric physics of the heating. Many papers have exploited this opportunity for precise measures, in this and other stars of its class — the "HW Vir stars".

## 7.2 Heating with Mass Transfer

Complications and asymmetries arise when there is mass transfer and an accretion disk. The supersoft binaries and their close cousins (the V Sge stars[31]) are then perhaps the best comparison class, since they have the basic requirements — a powerful source of heating and a donor star for possible irradiation. We have conducted long-term photometric campaigns on two such stars: QR Andromedae, a galactic supersoft binary (McGrath et al. 2002), and V617 Sgr, presently classified as a V Sge star (Steiner et al. 2006). Each has a very broad minimum ("eclipse") in its orbital light curve; and for each, we obtained ~400 hours of photometry over ~100 nights during 2003−2020. The CBA mean orbital light curves (averaged over ~20 orbits in the season of best coverage) of these stars are shown in the upper frames of Figures 5 and 6. There is a clear similarity to AA Dor, although the "eclipse" shapes and depths suggest that if they are truly eclipses, they must be *partial*. Actually, this seems to be the *typical* orbital waveform of both the supersofts and the V Sge stars — as exemplified by their prototypes (CAL 87 and V Sge; see Figure 4 of Schandl et al. 1997 and Figure 3 of Patterson et al. 1998). And comparison to Figure 1 shows a close resemblance to IM Nor and T Pyx, too.

---

[31] Both supersofts (van den Heuvel et al. 1992) and V Sge stars (Diaz & Steiner 1998, Patterson et al. 1998, Hachisu & Kato 2003) are very luminous mass-transfer binaries with WDs accreting at a high rate (~$10^{-7}$ $M_\odot$/yr). Early work suggested that the supersofts are quite distinct in their continuum spectra, showing an intense (and defining) 30–80 eV component, attributable to nuclear burning on the WD. More recent work shows that V Sge stars can be *transient* supersoft sources, and that both classes have high and low states in both visual and soft X-ray brightness. The optical spectra are also similar, with many high-excitation emission lines — quite uncharacteristic of ordinary CVs. So the distinctions between V Sge stars and supersofts have become much less convincing. Both class names will likely be replaced by something better, and we suspect that there is no fundamental difference.

The similarity extends also to a rapid change in orbital period. We measured the times of primary minimum in our data, and these times are given in Tables 2 and 3. For V617 Sgr, we added our 46 new timings to the 37 collected by Steiner et al. (2006), and the 3 presented by Shi et al. (2014). The composite O−C diagram is in the lower frame of Figure 5. The eclipse minima are tracked by the ephemeris

Min. light = HJD 2,446,878.773(2) + 0.207165513(2) $E$ + 5.5(2)×10$^{-11}$ $E^2$   (5).

The quadratic term corresponds to a rate of period increase $P/\dot{P}$ = 1.0×10$^6$ years.

For QR And, we combine all previously published (or extracted from published data) timings with our own, list them in Table 3, and reduce them to an O−C diagram in the lower frame of Figure 6. The eclipse minima are tracked by the ephemeris

Min. light = HJD 2,448,818.174(2) + 0.6604618(8) $E$ + 3.5(2)×10$^{-10}$ $E^2$   (6).

The quadratic term corresponds to a rate of period increase $P/\dot{P}$ = 1.7×10$^6$ years.

Both stars show obvious, rapid period increases. In our interpretation, all these phenomena arise from rapid mass-transfer to the WD, with consequent WD heating and irradiation of the (otherwise) innocent donor star.[32]

These seem to be good templates for understanding the light curves of IM Nor and T Pyx. For T Pyx, the term "eclipse" (implying a conjunction of the component stars) is supported by the phasing of the UKS radial-velocity data, and by the soft X-ray eclipse in eruption — at the phase of WD superior conjunction — reported by Tofflemire et al. (2013). No such evidence exists for IM Nor. But the resemblances to T Pyx are numerous. In addition to the eruption/decline similarities discussed by S10, there are the resemblances (in $P$, $\dot{P}$, and $M_V$) discussed above — all of which are unusual for classical novae — and the changes in orbital light curve as the stars fade from eruption. That's a lot of resemblance.

Simply put, IM Nor looks like T Pyx with a higher binary inclination and a higher interstellar absorption.

Of course, minimum light at superior conjunction of the WD does not guarantee that the dip is truly an eclipse (which implies a blockage of light). But one other recurrent nova, CI Aql, shows a similar orbital light curve, with a very wide "eclipse" when bright, and an eclipse of normal width and shape when quiescent (Schaefer 2011, Wilson & Honeycutt 2014). The wide eclipse in outburst and the narrow eclipse in quiescence are *in phase*. So the primary dips in all these stars probably do signify true geometrical eclipses of the bright component (the WD surrounded by a much larger structure — the accretion disk and/or a large "corona").

---

[32] Since both supersofts and V Sge stars show high and low states, it is possible that some prompt nuclear burning on the WD (not just accretion energy) also plays a role in irradiating the donor. And it is possible that additional very precise eclipse timings may show small changes in O−C which correlate with the on-off transitions (if the latter are not frequent).

Another dip in the orbital light curves occurs at or very near phase 0.5, when the donor is directly behind the WD (or luminous structures surrounding the WD). This suggests that the donor is also very luminous, although it is alternatively possible that the eclipsed light arises at the "hot-spot" where a mass-transfer stream strikes the periphery of a very large accretion disk. Apparently these are not ordinary CV donor stars!

It would be fascinating to learn what constraints on system geometry can be placed by these individual light curves, as attempted for CAL 87 by Schandl et al. (1997). But that's a project for another day — and another team.

### 7.3  Heating and Orbital Period Change

The heating of the secondary in AA Dor is obvious in the light curve, and this is confirmed by detailed measurement of day-side and night-side temperatures. No such analysis seems possible in the more complex geometry of IM Nor and (possibly) related stars.

In general, CVs are unpromising places for heating donor stars. In the absence of obstruction/shadowing, roughly 2% of the WD's light would fall on the secondary. Under most circumstances, this is greatly diminished, because the disk radiates not symmetrically, but *perpendicular* to the orbital plane, and because there may well be obstructions on the sight-line to the secondary (the height of the disk, the concave "flaring" of a theoretical disk, and any bright spot at the disk's edge). Such effects can greatly reduce the flux incident on the secondary. In addition, CV secondaries are usually *cool* stars, with plenty of neutral hydrogen in their atmospheres. So the opacity to UV/EUV radiation is very high, and little light can actually reach the photosphere (e.g. Hameury et al. 1986, King 1989). These are formidable obstacles, and probably explain why reflection effects are very rare in CV light curves.

During a classical-nova eruption, these problems disappear or are greatly mitigated. The obstructing disk is blown away, a shell is ejected with $\sim 10^{45}$ rrgs, and for at least a few months, the hot WD shines with $L \sim 10^{38}$ erg/s and $R > 1\ R_\odot$. A lot of flux (radiation and particles) must be incident on the secondary — vastly more than AA Dor's feeble secondary, which certainly has strong photospheric heating [18,000 K on the "dayside", according to Hoyer et al. (2015)]. Therefore, once the disk is blown away, it's plausible that the donor's hydrogen blanket may temporarily ionize and permit strong heating of the donor.

All that luminous energy deposited high in the secondary's atmosphere can result in a powerful wind from the secondary (Tout & Hall 1991; van Teeseling & King 1998, hereafter VTK; Knigge, King, & Patterson 2000, hereafter KKP). This can produce wind-driven-mass-transfer (WDMT), and its inner workings are as follows. A strong wind blows from the heated secondary, carrying away a lot of angular momentum (if the mass ratio is extreme). This contracts the binary dimensions and thereby throttles the secondary, which loses mass rapidly and therefore expands adiabatically, with $R \sim M^{-\frac{1}{3}}$. That transferred matter accretes onto the low-angular-momentum member of the binary (the WD) and thereby widens the binary (increases the period). As shown by VTK and KKP, orbital-period *increase* is expected from these effects if the donor is of low mass.

VTK invoked this WDMT model to account for supersoft binaries of short $P_{orb}$ (<6 hours). Such stars have high luminosity ($L_X \sim 10^{37}$ erg/s), and mostly in soft X-rays, which are ideal for absorption high in the donor's atmosphere, where a stellar wind is likely formed.

KKP used essentially this model to account for the large period increase in T Pyx. However, T Pyx has a lower luminosity ($\sim 10^{36}$ erg/s ) and, except in eruption,[33] is not detected in soft X-rays. As stressed by KKP, it is difficult to understand how accretion energy alone could irradiate the secondary with enough flux to power a wind at the required rate, which is approximately the accretion rate (VTK, KKP). Even for a massive (1.2 $M_\odot$) white dwarf, the accretion energy is still ~30× less than the nuclear energy. And because IM Nor and T Pyx show nova eruptions, it seems likely that they do not burn H between eruptions. So how can they mimic the supersoft binaries?

We leave this as an unsolved problem. One possible answer could be found in the question: how frequently are IM Nor and T Pyx "in eruption" (burning H)? They are classified as slow novae (Caleo & Shore 2015, SSH), and erupt every ~40 years. It seems possible that they might spend ~3% of their time burning H — and likely at a rate much higher than the well-known supersofts, since the novae reach an Eddington luminosity, eject matter, and — at least for T Pyx — exhibit a huge *dP/dt* in outburst (at least 30× greater than at quiescence; see Figure 6 of P17). If the irradiated donor star retains some memory (viz., *energy*) from this violent event, then maybe what matters most for WDMT is the *time-averaged* irradiation.

But what about the ~1 year interval of actual eruption, when the WD is 100−1000× more luminous? That irradiation should produce a much greater wind from the donor, therefore much greater mass transfer, and therefore much greater period increase. That would compete with, and possibly dominate, the effect of simple mass-loss from the WD, carrying away negligible angular momentum (assumed in P17) since the mass ratio $M_2/M_1$ is very low.

So we now think our P17 conclusion of $10^{-4}$ $M_\odot$ ejected from the WD in T Pyx was premature. When a WD explodes, it's easy, but hazardous, to think that whatever is in its vicinity is just going along for the ride.

Sudden mass ejection would lead to a sudden change in period, and we have studied our 2011 timing data for this. But Figure 4 of P17 shows that the orbital modulation after outburst was immeasurably small, only growing to 0.01 mag full amplitude ~200 days after outburst. And the accumulated phase difference (between the scheduled pre-outburst and post-outburst eclipses) is only 0.1 cycles. Inspection of Figure 7 in P17 shows that this is roughly the accuracy of our previous conclusion that the period increase was sudden and consistent with the day (or interval) of eruption ("day 120±90"). To this we now add "also consistent with a gradual Δ*P* due to a short-lived but very large enhancement of the WDMT".

---

[33] Tofflemire et al. (2013) give details of the detections and nondetections in eruption. Because much of the X-ray flux is in emission lines, and because interstellar absorption and binary inclination are not well known, the constraints are not strong.

In the latter interpretation, the donor star is always chock full of energy from hundreds of previous eruptions and their 30-year aftermaths; and the resultant wind drives high mass-transfer and the long-term $\dot{P}$. In the April 2011 eruption, the WD merely made another large deposit to the donor's energy supply. Some of that new energy powered a greatly increased and short-lived $\dot{P}$, and some was stored to power the WDMT during the next 40 years. This achieves a better economy of hypothesis. What's good for the goose (quiescent $\dot{P}$) may well be good for the gander (eruption $\dot{P}$).

### 7.4 The Donor Star, Revisited

An important and surprising lesson from these considerations (overall waveform, secondary eclipses, period change, etc.) is that the donor in these short-period binaries is probably far from the cool, faint, unimportant star envisioned by most current theories of CV evolution. Tormented by radiation from its partner, it may well be hot, bright, and losing mass copiously to both its partner and its own stellar wind. Except during the nova eruption, it is probably the powerhouse of the binary.

### 8 CATACLYSMIC-VARIABLE ESCHATOLOGY

Theory suggests that cataclysmic variables, late in life, reach a minimum orbital period when the secondary begins to expand in reaction to mass loss. If the driver of evolution is angular momentum loss due to gravitational radiation (GR), that minimum period should be near 70 minutes (e.g., Rappaport et al. 1983, Barker & Kolb 2003). Observations are *roughly* in accord with this expectation, although a somewhat stronger driver [e.g., $\dot{J} = 2.5\ \dot{J}_{GR}$; Knigge et al. 2011, Figure 6] would improve the fit, moving the theoretical minimum period close to the observed number (~80 minutes). This would also make the stars systematically brighter than they would be from GR alone, make the donors at a given $P_{orb}$ less massive than from GR alone (as observed, see Figure 6 of P11), and make the lifetimes shorter than they would be from GR alone.

What would be that extra driver? Well, theoretical models usually assume that nova eruptions are insignificant short-lived blips in the overall evolution story. This may be warranted for most novae, which have "long" orbital periods (> 3 hr) and indeed do return to near-quiescence after ~100 years. But all known short-$P_{orb}$ novae are still ~4 magnitudes above their proper quiescence — at least 100, and in one good candidate (BK Lyn) 2000, years after eruption. The estimates of P13 suggested that the back-reaction of all this extra light on the donors might keep the mass-transfer rate high and thereby shorten the lifetimes of these low-mass donors. "Shorten the lifetime" depletes the census of such stars, which might solve the long-standing difficulty in understanding why 99% of all CVs are not of short $P_{orb}$ (because to account for the 100× difference in accretion rates, magnetic braking needs to be ~100× stronger than GR).

IM Nor and T Pyx may be the extreme examples of this process. As described by KKP, they may have left the CV evolution track altogether, and joined an eruption-quiescence cycle powered by WDMT, which would only end with the donor evaporated or nearly so.

But the vast majority of short-period CVs contain cool (~15,000 K) WDs and very cool (~3,000 K) donors. Such stars are presumed to be the descendants (as well as ancestors) of most short-$P_{orb}$ novae. We do not understand what parameter can select the T Pyx stars for such vastly different evolution. Here's our best guess (just to be sporting about it). The WDMT mechanism selects for low $M_2$ because that carries away more angular momentum, and more importantly because such a star will expand on being severely heated. It also selects for high $M_1$, because massive WDs are expected to erupt much more frequently. So a very sharp dependence on mass ratio $q = M_2/M_1$ might be the answer. Future observational estimates of $q$ for the 7 known short-period old novae might test this.

Because nova recurrence time depends so sharply on WD mass, nova eruptions on a massive WD may start the binary on a quick path to destruction. The donor is heated with each successive eruption, fails to cool sufficiently to join its cousins at the lower left of Figure 3, and thus becomes more strongly heated (outside-in) with each successive eruption. Thus a strong wind is maintained between eruptions, which keeps the WDMT and thus the orbital-period change at a high level between eruptions, even though the nuclear burning is turned off 97% of the time. There's a big difference between 3% and 0.001%.

## 9  1866 AND ALL THAT

In contrast to this "evolution model", one could also consider a "seizure model", which depicts these recurrent-nova events as basically a series of short-lived seizures in the aftermath of some other (postulated) larger event — which could even be a (somehow different) classical-nova eruption. For T Pyx, the motivator for this model, Schaefer (2005), Schaefer et al. (2010, hereafter SPS), and Shara et al. (2016) have advocated this view — suggesting that a putative (unobserved) classical-nova eruption around the year 1866 triggered the sequence of later outbursts.

In support of this, Schaefer (2005) points out that the interval between successive eruptions has been lengthening somewhat since 1890, the first observed eruption, and that the post-eruption brightness level has faded significantly and steadily since then. To our eye, the supporting data looked impressive. And our data, based on ~500 nights of observation, lends some support to this: the average *V* magnitude of T Pyx during 1996−2009 was ~15.4, smoothly declining to ~15.85 during 2018−2020. Although the SPS 2010 prediction of future behavior ("no eruptions for the foreseeable future") was promptly slapped down by T Pyx itself in its 2011 eruption, the main point remains: there are grounds for suspecting that the recent frenzy of rapid eruptions may be a short-lived episode (a few hundred years).

If true, the "1866 theory" downgrades the credentials of T Pyx, and probably IM Nor, as an avatar of binary evolution. On the other hand, the 2011 T Pyx eruption was probably the most extensively studied nova in history, and basically "checked all the boxes" as a true classical nova event: light curve, luminosity, mass ejected, radio emission, stages of spectral development, orbital-period change, etc    So the 1866 theory probably needs to confront the question: what was this event which launched a series of rapid, powerful, seemingly normal nova explosions? If it was a simple classical nova, why have the ~100 other classical novae with $P_{orb}$<0.5 days not triggered similar repetitions?

## 10  WHY NO SUPERHUMPS?

In our earlier study of ~200 novalike variables and dwarf novae (Patterson et al. 2005, especially their Figure 8), we found a simple rule for the sustained "high states" ($M_V$<+6) of nonmagnetic CVs: common superhumps are present in all stars with $P_{orb}$<3.3 hr, and absent in all stars with $P_{orb}$>4 hr. Most theoretical work (*e.g.*, Lubow 1991) considers the mass ratio $q = M_2/M_1$ to be the controlling physical parameter, with $q$<0.35 as the requirement for superhumps.

T Pyx and IM Nor are recurrent novae (implying high $M_1$) of short $P_{orb}$ (implying low $M_2$), and are plenty luminous ($M_V$~+2). But their light curves lack superhumps; the limit for IM Nor is fairly coarse (<0.03 mag), but that of T Pyx is very stringent (<0.003 mag). We have also found no superhumps in any of the V Sge or supersoft binaries we have observed, although they have longer orbital periods and may possibly flunk the mass-ratio requirement.

So, at least in the two stars highlighted in this paper, the absence of superhumps seems puzzling.

Superhumps are an accretion-disk phenomenon — essentially arising from prograde apsidal precession of the disk. Is it possible that these stars do not have accretion disks?

That seems unlikely. In T Pyx the double-peaked spectral lines certainly suggest a disk origin (Uthas, Knigge, & Steeghs 2010); and for all these stars, the optical and X-ray "eclipses" suggest a very broad light source centered on the WD. With no evidence for magnetic channeling, an accretion disk is the natural way for accreting gas to reach the WD

We leave this question without solution. It is possible that superhumps are restricted to stars with a somewhat narrow range of accretion rates — favoring $\dot{M}$ near ~$10^{-8}$ $M_\odot$/yr, but intolerant of large departures from this standard. Another possibility is that very energetic photons (from H-burning on the white dwarf) may change the disk structure in some way that suppresses the growth of the instability which makes superhumps.

Superhumps were a shocking discovery of the 1970s, and not understood for ~15 years; so they probably still contain a mystery or two.

## 11  SUMMARY

1. We report a campaign of time-series photometry of IM Nor, a classical nova with a recurrence time of 82/$N$ years, where $N$ could be 1, 2, 3, or 4. During the aftermath (2003−2020) of the most recent eruption, the brightness declined from $V$=17.5 to ~18.7.

2. We compare IM Nor with T Pyx, the only other short-$P_{orb}$ recurrent nova currently known. They show similar eruption light curves, and orbital light curves which are similar at every stage (near maximum, at "quiescence", and on the decline to quiescence). We estimate that IM Nor's quiescent Mv, corrected for interstellar absorption, is roughly +2.8 — about 3 magnitudes brighter than typical novae. If this light comes from accretion, the accretion rate should be near $10^{-7}$ $M_\odot$/yr.

3. The orbital light curve shows shallow, broad "eclipses", similar to that of T Pyx and several supersoft binaries. But the eclipse's great width and shallow depth show that the component "stars" (sources of light) in these binaries must be of a size comparable to their Roche lobes, and the low-mass donor star must be strongly heated.

4. Times of mid-eclipse show that the orbital period is increasing on a timescale of $1.3 \times 10^6$ years. Interpreted as an effect of mass transfer with total mass and angular momentum conserved, this implies a mass-transfer rate near $10^{-7}$ $M_\odot$/yr. This is also roughly the rate needed to power recurrent-nova outbursts.

5. With $P_{orb}$ = 2.46 hours, IM Nor must have a low-mass secondary (<0.25 $M_\odot$). Since that secondary appears to be losing mass at a hefty rate, IM Nor may only last for ~2 million years before completely cannibalizing its secondary. Mass loss during the actual nova eruption would further hasten its demise. So rapid a self-destruction may explain the rarity of such stars.

6. Similar campaigns on QR And (a supersoft binary) and V617 Sgr (a V Sge star) yield similar mean light curves and rates of $P_{orb}$ increase. A single mechanism may power evolution in all these hot, luminous stars with $P_{orb}$< 1 day, and we identify that as wind-driven mass transfer (WDMT). This does, however, impugn the "conservative mass transfer" assumption, since there must be significant angular-momentum loss in WDMT. A more thorough investigation of this is warranted.

7. There is also a lingering difficulty in understanding how a source presumed to be powered by gravitational energy can closely mimic a nuclear-powered source. A possible answer may lie in the ***time-averaged*** high-energy radiation incident on the donor star.

8. In all these luminous binaries with increasing $P_{orb}$, the donor is probably not the cool, faint, innocent star envisioned by most theories of CV evolution. The dynamics of the binary, plus the mere existence of secondary eclipses, suggest strong irradiation by the hot WD, which can be sufficient to drive a wind which stimulates a long-lived episode of high mass transfer.

9. To any student of CV evolution, it is startling to see short-$P_{orb}$ CVs with such high mass-transfer rates in a state nominally called "quiescence". Unless something like the 1866 theory is correct, IM Nor and T Pyx offer an opportunity, sought for many years, to kill off short-period CVs (by evaporating the donor).

10. On the other hand, we have not explained why IM Nor and T Pyx appear to be different from the other five old novae of short $P_{orb}$. Three of the five erupted from states of very low luminosity (CP Pup, V1974 Cyg. GQ Mus; Schaefer & Collazzi 2010), so we presume these five to be different — and we do not yet understand that difference. Long-term studies of $P_{orb}$ for any of the five would be helpful.

11. It is possible that T Pyx's apparently sudden increase in orbital period during the 2011 outburst arose mainly from a sudden increase in the WDMT, rather than a sudden expulsion of ~$10^{-4}$ $M_\odot$ from the white dwarf. If so, that undermines the P17 conclusion that the white dwarf in T Pyx is gradually losing mass, and reopens the possibility that T Pyx stars may eventually explode as supernovae.

12. Many studies have remarked on the puzzling rarity of old CVs — stars which have evolved past "period bounce" at ~80 minutes (Patterson 1998, Barker & Kolb 2003, Littlefair et al. 2006, P11, Knigge et al. 2011, Pala et al. 2018). Most assume evolution in or near that regime to be "quiet" — dominated by the inevitable but very slow grind of gravitational radiation. Perhaps it is time to consider unquiet paths of evolution. Intensive study of the other five short-period novae might be a good start.

Long-term study requires long-term commitment, organization, and funding. For the latter, we are grateful for "seed money" from the Research Corporation and the Mount Cuba Astronomical Foundation, and steady support from the National Science Foundation (most recently AST-1908582) and NASA (HST-GO-15454.002-A). We thank Dave Skillman and Dave Harvey for starting us down this road 35 years ago. We thank Brad Schaefer, Jeno Sokoloski, and Koji Mukai for numerous conversations and dogged pursuit of the matters discussed here. Most of all, we thank the AAVSO, BAA, RASNZ, and SAS (Society for Astronomical Sciences) for providing the social and organizational structure needed to keep humans attentive to such arcane matters for many decades!

**FIGURE CAPTIONS**

Figure 1. *Upper frames:* the orbital light curves of IM Nor in 2003 and 2017, averaged over ~25 orbits. *Lower frames:* the orbital light curves of T Pyx in 2012 and 2017, averaged over ~40 orbits. The light curves and trends (eclipse depths increasing with time after eruption) are similar. Additional T Pyx orbital light curves are given in Figure x of P17.

Figure 2. O−C diagram of the IM Nor eclipse timings, with respect to a test period of 0.10263317 d. The parabolic fit indicates steady period increase, given by Eq. (1).

Figure 3. Empirical plot of <$M_V$> versus $P_{orb}$ for CVs. Crosses are "normal" CVs (dwarf novae and novalike variables). Dots are historical novae at "quiescence" (25-150 years after eruption). IM Nor and T Pyx, identified by name, are far above their colleagues of similar $P_{orb}$.

Figure 4. CBA orbital light curve of AA Dor, a highly inclined and *detached* binary in which a low-mass secondary is heated by the hot primary's radiation. This well-studied star serves as a template for understanding the orbital light curves of close mass-transfer binaries with reflection effects.

Figure 5. *Upper frame:* CBA orbital light curve of V617 Sgr, a V Sge star. *Lower frame:* O−C diagram for the primary minimum in its eclipse cycle — showing a rapid period increase.

Figure 6. *Upper frame:* CBA orbital light curve of QR And, a supersoft binary. *Lower frame:* O−C diagram for the primary minimum in its eclipse cycle — showing a rapid period increase.

**TABLE 1**

**Mid-Eclipse Times for IM Nor**

(HJD 2,450,000+)

| | | | |
|---|---|---|---|
| 2696.5260 | 5030.2958 | 7139.6185 | 7871.0871 |
| 2754.5137 | 6341.4375 | 7140.2322 | 7872.3198 |
| 2782.4293 | 6347.4915 | 7141.1559 | 7874.1660 |
| 2842.2620 | 6362.4751 | 7148.2393 | 8221.0711 |
| 3165.2500 | 6363.5026 | 7149.2645 | 8228.1533 |
| 3168.8433 | 6387.5179 | 7150.0873 | 8229.1791 |
| 3172.6390 | 6387.6209 | 7150.2924 | 8230.2050 |
| 3177.5680 | 6394.4976 | 7465.0664 | 8232.0514 |
| 3179.6180 | 6738.5248 | 7469.0709 | 8578.0330 |
| 3180.5409 | 7134.4863 | 7472.1510 | 8967.9383 |
| 3182.5953 | 7137.2563 | 7858.2564 | 8976.9698 |
| 5015.3115 | 7139.4135 | 7867.0843 | |
| 5026.2946 | 7139.5151 | 7870.9842 | |

**TABLE 2**

**New Mid-Eclipse Times for V617 Sgr**

(HJD 2,400,000+)

| | | | |
|---|---|---|---|
| 53891.6035 | 53922.4706 | 56177.9306 | 57156.1964 |
| 53911.0786 | 53923.3004 | 56181.9693 | 57157.2314 |
| 53911.2845 | 53923.5055 | 56182.9028 | 57158.0604 |
| 53911.4906 | 56143.3323 | 56187.8756 | 57515.2210 |
| 53911.9066 | 56150.3820 | 56460.5156 | 57552.0966 |
| 53912.3915 | 56162.8050 | 56463.6215 | 57905.3213 |
| 53918.3266 | 56163.0140 | 56480.4044 | 57906.1523 |
| 53918.5354 | 56167.9891 | 56823.6866 | 57907.1873 |
| 53920.4035 | 56169.0244 | 56824.7206 | 58704.5885 |
| 53920.6066 | 56169.8528 | 56825.7566 | 59061.5463 |
| 53921.4335 | 56174.8255 | 56827.6224 | 59064.8526 |
| 53922.2634 | 56175.8579 | 56830.7274 | |

**TABLE 3**

**All Mid-Eclipse Times for QR And**

(HJD 2,400,000+)

| HJD | Source | HJD | Source | HJD | Source |
|---|---|---|---|---|---|
| 35799.247 | 1 | 50434.323 | 2 | 57679.649 | 4 |
| 37260.190 | 2 | 51753.922 | 3 | 58009.8866 | 4 |
| 42983.693 | 1 | 56586.5743 | 4 | 58011.8691 | 4 |
| 48093.645 | 1 | 56592.5042 | 4 | 58012.533 | 4 |
| 48887.509 | 3 | 56594.5052 | 4 | 58013.8441 | 4 |
| 49743.485 | 1 | 56602.3970 | 4 | 58015.8396 | 4 |
| 49987.846 | 3 | 56908.8955 | 4 | 58017.8317 | 4 |
| 50073.715 | 4 | 56916.8096 | 4 | 58018.4780 | 4 |
| 50369.590 | 2 | 56927.3770 | 4 | 58023.7574 | 4 |
| 50379.504 | 2 | 57260.1900 | 4 | 58758.8644 | 4 |
| 50391.385 | 2 | 57671.7327 | 4 | 58760.8500 | 4 |
| 50397.353 | 2 | 57673.7058 | 4 | 58764.8130 | 4 |
| 50430.361 | 2 | 57675.6956 | 4 | | |

1 = Greiner & Wenzel 1995
2 = Deufel et al. 1999
3 = McGrath et al. 2002
4 = this work

**FIGURE 1**

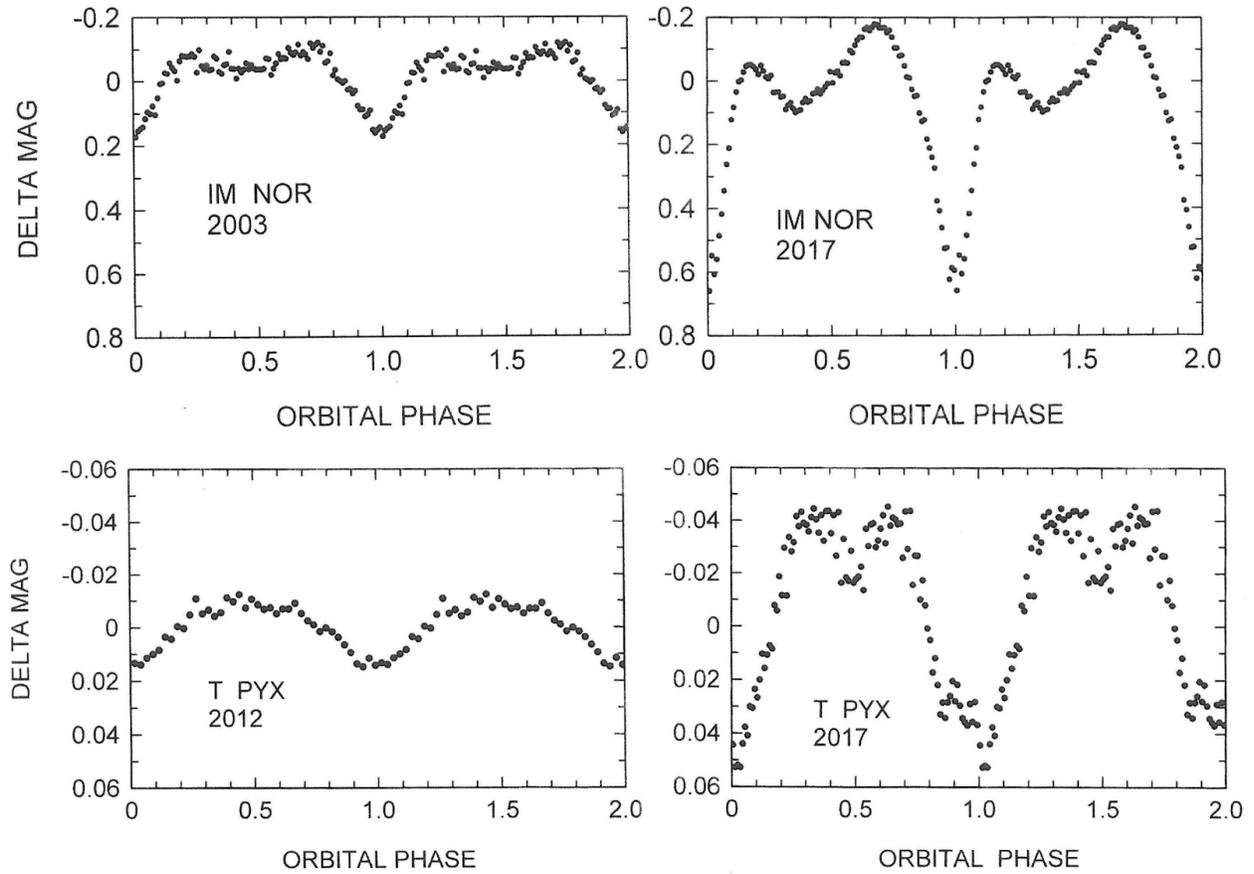

Figure 1. *Upper frames:* the orbital light curves of IM Nor in 2003 and 2017, averaged over ~25 orbits. *Lower frames:* the orbital light curves of T Pyx in 2012 and 2017, averaged over ~40 orbits. The light curves and trends (eclipse depths increasing with time after eruption) are similar. Additional T Pyx orbital light curves are given in Figure x of P17.

**FIGURE 2**

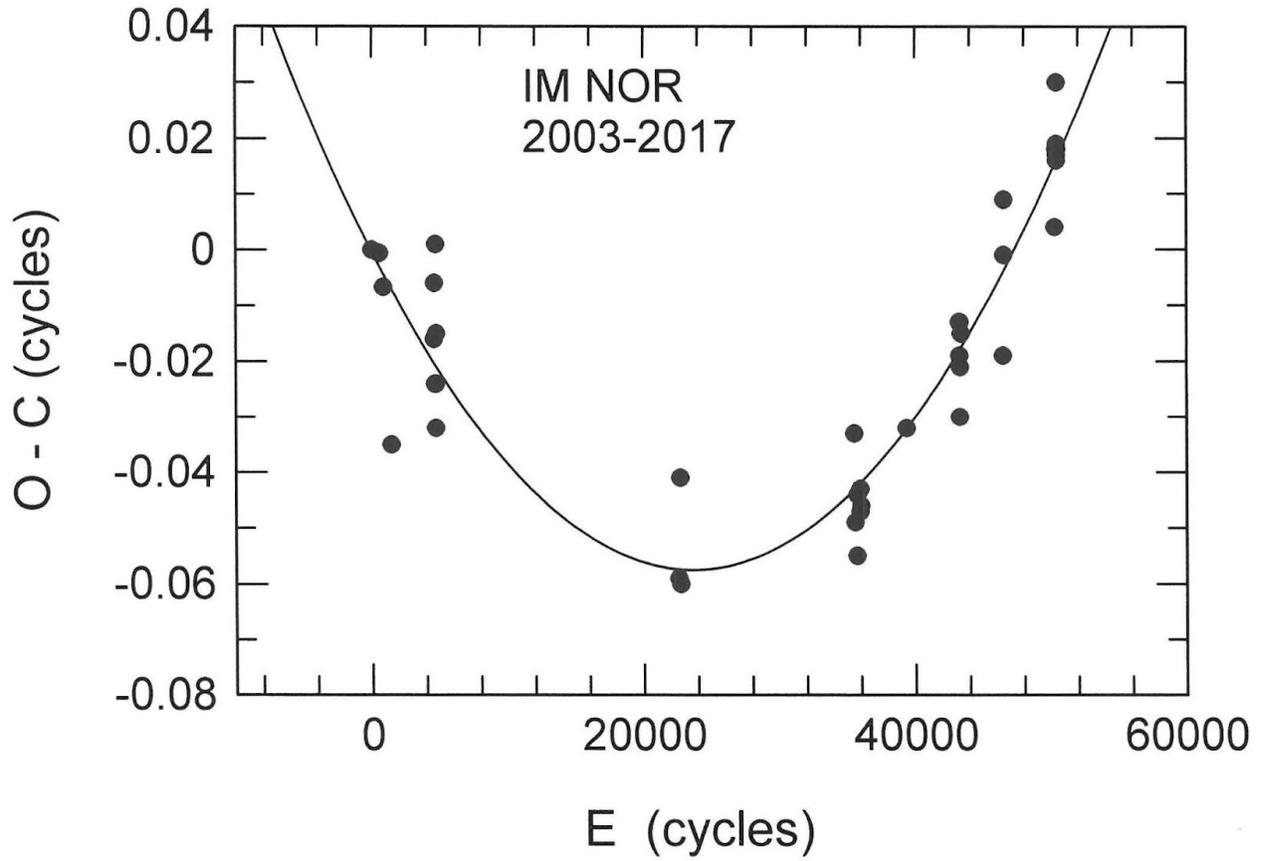

Figure 2. O−C diagram of the IM Nor eclipse timings, with respect to a test period of 0.10263317 d. The parabolic fit indicates steady period increase, given by Eq. (1).

**FIGURE 3**

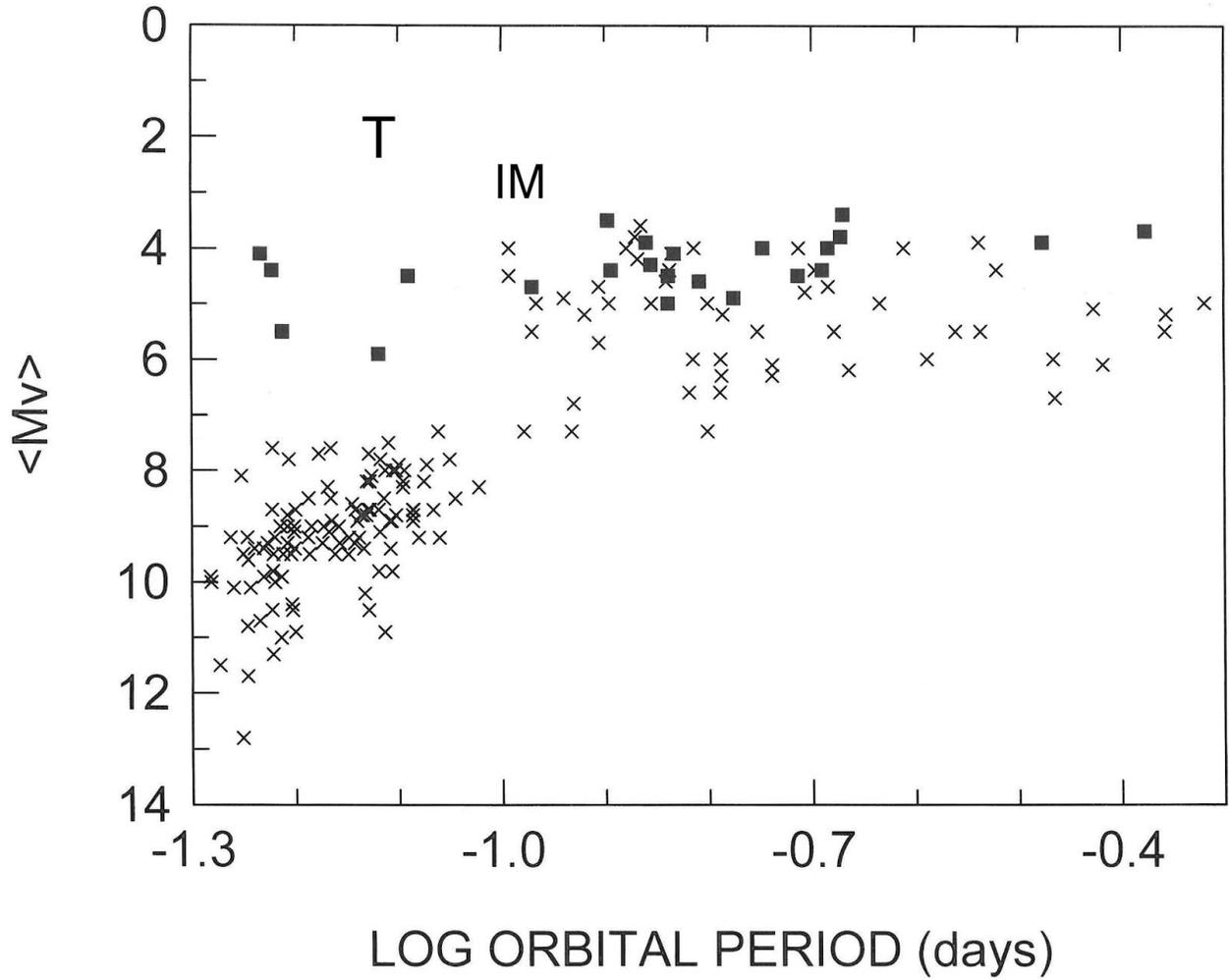

Figure 3. Empirical plot of $<M_V>$ versus $P_{orb}$ for CVs. Crosses are "normal" CVs (dwarf novae and novalike variables). Dots are historical novae at "quiescence" (25-150 years after eruption). IM Nor and T Pyx, identified by name, are far above their colleagues of similar $P_{orb}$.

**FIGURE 4**

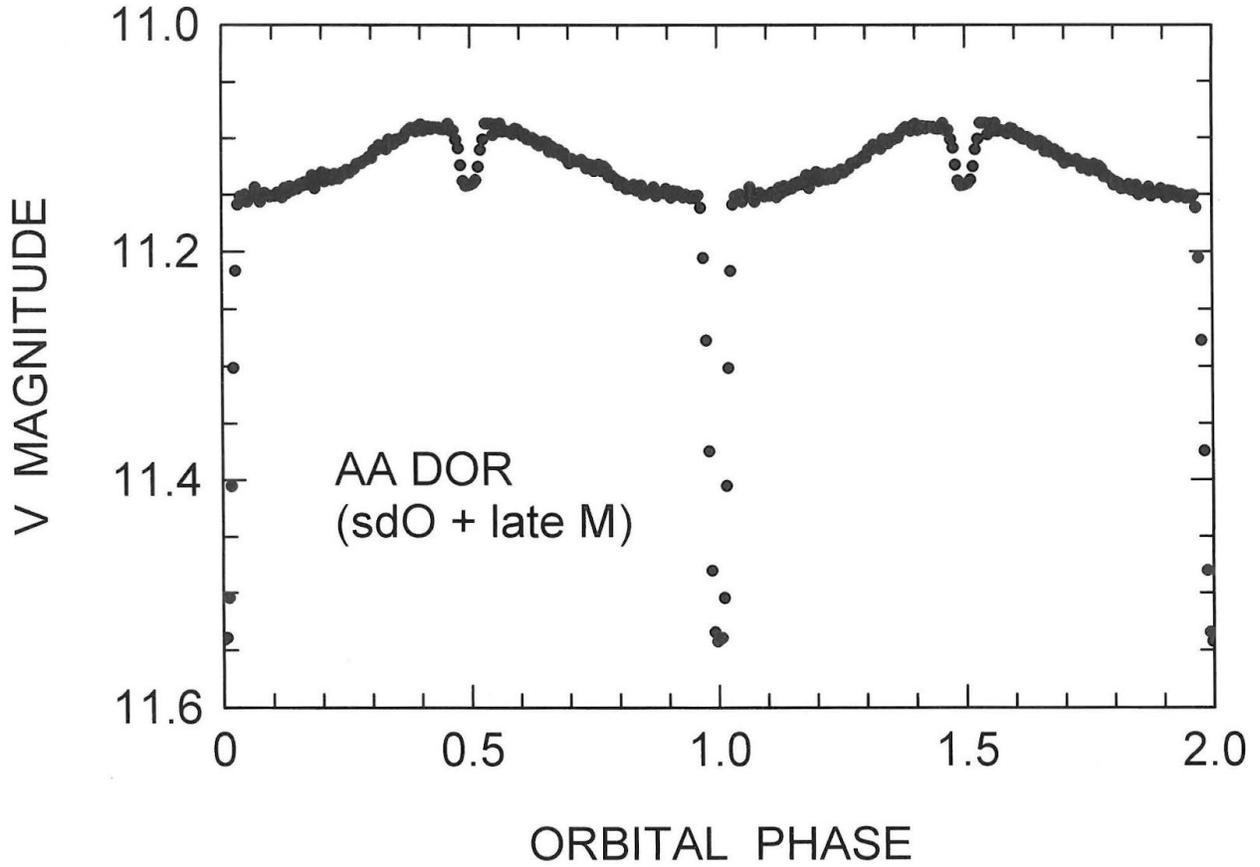

Figure 4. CBA orbital light curve of AA Dor, a highly inclined and *detached* binary in which a low-mass secondary is heated by the hot primary's radiation. This well-studied star serves as a template for understanding the orbital light curves of close mass-transfer binaries with reflection effects.

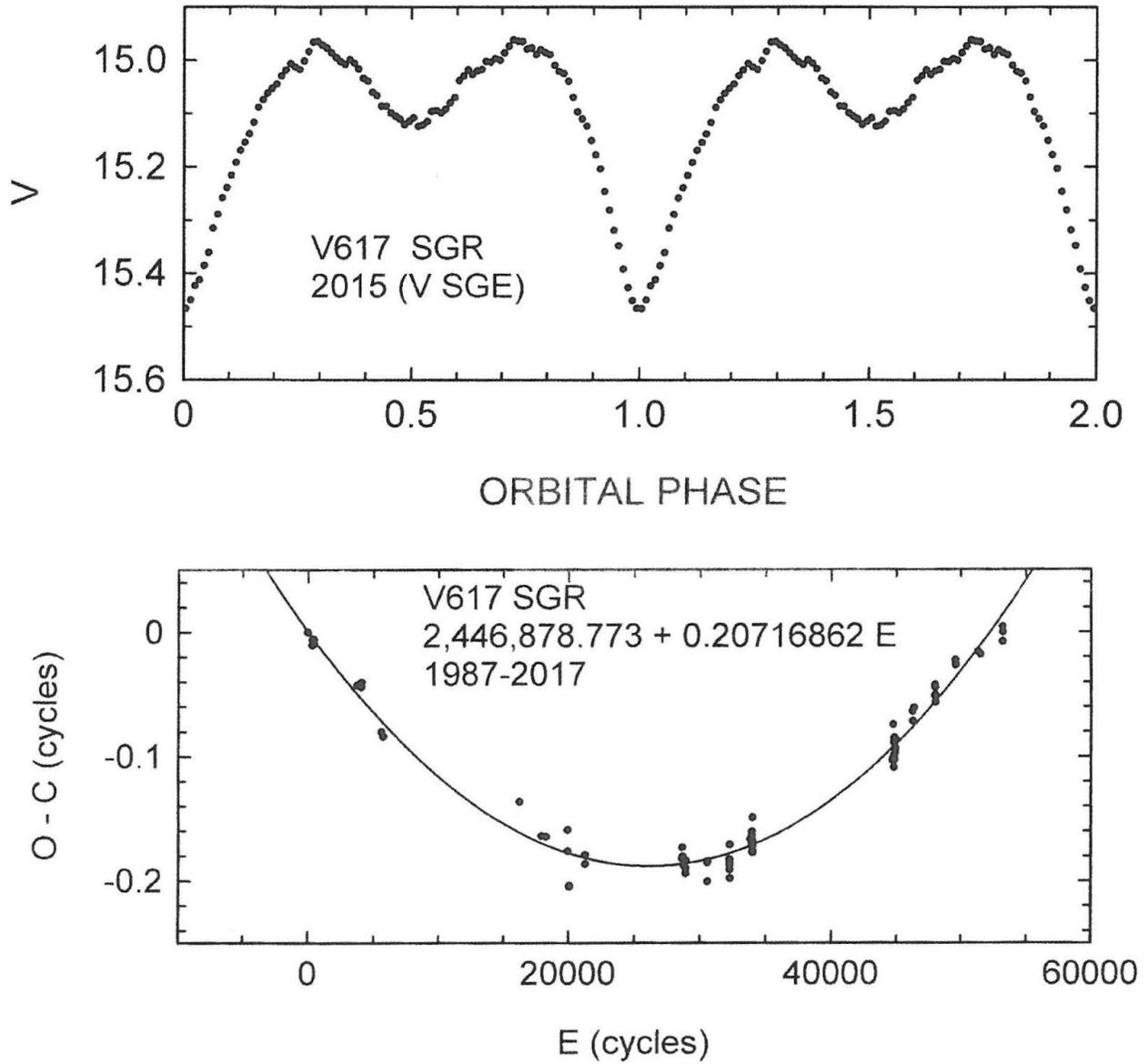

Figure 5. *Upper frame:* CBA orbital light curve of V617 Sgr, a V Sge star. *Lower frame:* O−C diagram for the primary minimum in its eclipse cycle — showing a rapid period increase.

**FIGURE 6**

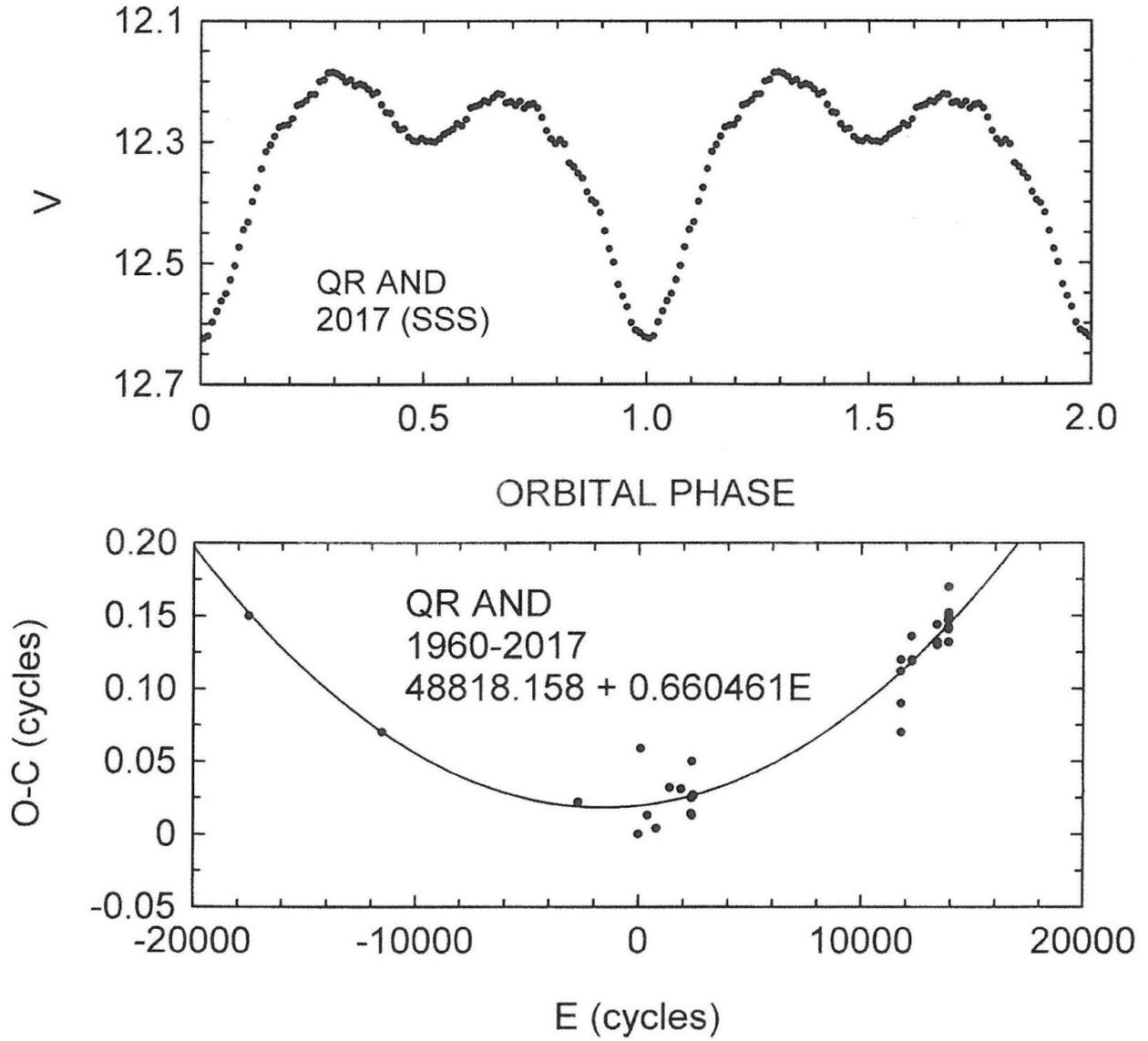

Figure 6. *Upper frame:* CBA orbital light curve of QR And, a supersoft binary. *Lower frame:* O−C diagram for the primary minimum in its eclipse cycle — showing a rapid period increase.